\documentclass[useAMS,usenatbib,usegraphicx]{mn2e}
\usepackage{color}

\title[The 11.2$\,\mu$m emission of PAHs in astrophysical objects]{The 11.2$\,\mu$m emission of PAHs in astrophysical objects}

\author[A. Candian and P.J. Sarre]{A. Candian$^{1,2}$\thanks{
 candian@strw.leidenuniv.nl} and P.J. Sarre$^{2}$\thanks{Peter.Sarre@nottingham.ac.uk} \\
$^1$Leiden Observatory, Niels Bohrweg 2, 2333 CA Leiden, The Netherlands.\\
$^2$School of Chemistry, The University of Nottingham, University Park, Nottingham NG7 2RD, U.K.\\}

\begin{document}

\date{Accepted. Received.}

\pagerange{\pageref{firstpage}--\pageref{lastpage}} \pubyear{2015}

\maketitle

\label{firstpage}

\begin{abstract}
The 11.2$\,\umu$m emission band belongs to the family of the `Unidentified' Infrared (UIR) emission bands seen
in many astronomical environments. In this work we present a theoretical interpretation of the band characteristics and profile variation for a number of astrophysical sources in which the carriers are subject to a range of physical conditions.  The results of Density Functional Theory (DFT) calculations for
the solo out-of-plane (OOP) vibrational bending modes of large polycyclic aromatic hydrocarbon (PAH) molecules are used as input for a detailed emission model
which includes the temperature and mass dependence of PAH band wavelength, and a PAH mass distribution that varies with object. Comparison of the model
with astronomical spectra indicates that the 11.2$\,\umu$m band asymmetry and profile variation can be explained principally in terms of the mass
distribution of neutral PAHs with a small contribution from anharmonic effects.
\end{abstract}

\begin{keywords}
molecular processes -- ISM: molecules -- ISM: lines and bands
\end{keywords}

\section{Introduction}

The discovery by \citet{Gill73} of an emission feature near 11.2$\,\umu$m in spectra of planetary nebulae opened a new era in the study of interstellar
matter.  It is now recognised as one of the strongest of the `unidentified infrared' (UIR) or `aromatic infrared' (AIB) bands that
lie between 3 and 20$\,\umu$m and which are generally attributed to vibrational transitions of polycyclic aromatic hydrocarbon (PAH) molecules
\citep[for a review see][and references therein]{Tiel08}. However, in no case has it proved possible to identify a specific PAH molecule.  This is
unfortunate because knowledge of PAH shape, size distribution and degree of hydrogenation and/or ionisation would allow much greater exploitation of the
spectra as probes of astrophysical conditions and processes. The advent of infrared (IR) satellites including ISO, Spitzer \citep{spitzer} and AKARI \citep{akari}
has revealed the richness of the AIB
spectrum and also significant differences in the profiles of individual bands \citep[\emph{e.g.}][]{Vand04,Bern07,Rose11,Rose12, Boer12, Boer13}. Interpretation of these
variations in terms of the physical and chemical properties of PAHs and the astronomical objects in which they are found could assist  in the
identification of PAH sub-groups, and thus narrow down the number of possible PAHs as carriers of AIBs.

In this paper we focus on the 11.2 $\,\umu$m feature, which is one of the most distinctive bands in the AIB spectrum.  It has been assigned to the C-H
out-of-plane (OOP) bending mode of solo-containing neutral PAHs \citep{Hony01,Vand04,Baus08, Ricc12}, although their precise molecular shapes and size distribution
has not been established.  Weaker emission features around 11.0~$\,\umu$m are generally attributed to the same type of transition in PAH cations
\citep{Sloa99,Hudg99,Boer13}. We present here an emission model based on Density Functional Theory (DFT) calculations of vibrational transitions
and intensities for a set of PAH molecules. The model considers the emission process following optical/UV excitation using the relevant stellar spectral
energy distribution (SED) and includes the temperature dependence of the emission wavelength (band position) as the PAH molecule cools through emission of
IR photons.  The results are used to explore the variation of the 11.2$\,\umu$m profile in a range of astronomical objects with particular reference to the
influence of the PAH mass distribution.

The paper is arranged as follows. In Section~\ref{sec:char} the characteristics of the 11.2$\,\umu$m band are reviewed. The theoretical approach and the results of a series of DFT calculations are presented in Section\,\ref{sec:calc} and the emission model is described in Section\,\ref{sec:model}.  Sections\,\ref{sec:res} and \ref{sec:dis} contain the results and discussion of their astrophysical implications. A possible contribution from acenes to the 11.0$\,\umu$m feature in the Red Rectangle is  also discussed.

 \section{Characteristics of the 11.2$\,\umu$\MakeLowercase{m} band}
 \label{sec:char}

 \subsection{Classification of the 11.2$\,\umu$\MakeLowercase{m} band}
\begin{figure}
\includegraphics[scale=0.58]{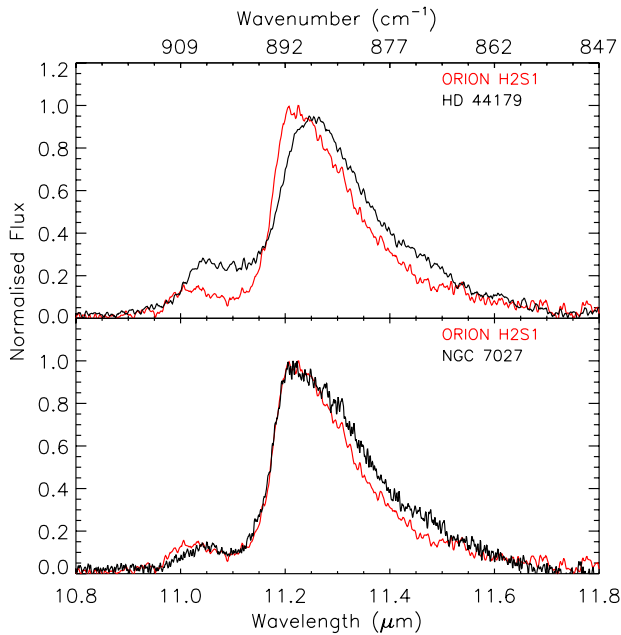}
\caption{Comparison of the 11.2$\,\mu$m profiles of class \textbf{A$_{11.2}$} (Orion H2S1) -- in red (thin line), \textbf{B$_{11.2}$} (HD~44179 - the Red Rectangle) and \textbf{A(B)$_{11.2}$} (NGC~7027).  These ISO  spectra are continuum-subtracted and normalised to the peak intensity.}
 \label{11.2_cl}
 \end{figure}

The 11.2$\,\umu$m feature has an asymmetric shape with a steep blue side and an extended red tail \citep{Roch89,Witt89}. Following analysis of ISO
observations of objects in the Galaxy, \cite{Vand04} proposed two classes of 11.2$\,\umu$m sources (\textbf{A$_{11.2}$} and \textbf{B$_{11.2}$}) and
introduced a description using a \textit{tail-to-top} ratio, \emph{i.e.} the strength of the red wing relative to the peak intensity of the overall profile
with maximum near 11.2$\,\umu$m.  \cite{Sloa07} introduced a third class, \textbf{C$_{11.2}$}. The class characteristics, illustrated with examples in Fig.
1, are:

\begin{description}
           \item [\textbf{A$_{11.2}$}] - the most common, with a peak wavelength ranging between 11.20 and 11.24$\,\umu$m and a relatively short red
               tail. Sources belonging to this class have a low value of \emph{tail-to-top} ratio and comprise various types of interstellar matter -
               reflection nebulae, H\,$\textsc{ii}$ regions and the general interstellar medium.
           \item[\textbf{B$_{11.2}$}] - less common, with a peak wavelength of 11.25 $\,\umu$m and a long tail. Objects in this class, such as the Red
               Rectangle (HD~44179), have a high value of \emph{tail-to-top} ratio and are usually associated with circumstellar matter.
            \item[\textbf{C$_{11.2}$}] - only a few examples, with peak wavelength ranging between 11.35 and 11.40$\,\umu$m and an approximately
                symmetric shape. Bands belonging to this class are found in carbon-rich objects.

\end{description}
Some objects (mostly planetary nebulae) are classified as class \textbf{A(B)$_{11.2}$} and have a mixture of the characteristics of the A and B classes, with a peak wavelength of class A$_{11.2}$ and a \emph{tail-to-top} ratio of class B$_{11.2}$ objects (see for example BD+30$^{\circ}$ 3639 in
Fig.\,\ref{11.2_cl}). Also, within the single reflection nebula NGC~7023, variation in the 11.2$\,\umu$m spectra from type A to A(B) has been found \citep{Boer13}. The A/B-type classification has also been applied to other spectral features in the 3$\,\umu$m and 6-9$\,\umu$m regions.

Based on a study of mid-IR \emph{Spitzer} spectra of carbon-rich post-AGB stars in the LMC, \citet{Mats2014} have introduced separate classifications for the 6-9$\,\umu$m and 10-14$\,\umu$m regions, where the spectra for the latter region are classified as $\alpha,\beta,\gamma$ and $\delta$.  As the spectra we consider here are high-resolution data from ISO-SWS \citep{degr96}, we have elected to use the \textbf{A$_{11.2}$}, \textbf{B$_{11.2}$} and \textbf{A(B)$_{11.2}$} classification of \cite{Vand04}.

\subsection{Asymmetry and peak wavelength of the 11.2$\,\umu$\MakeLowercase{m} band}
Interpretation of the shape and the variation in the profile of the 11.2$\,\umu$m band is a significant challenge which requires an explanation of (a) the
steep short-wavelength side, (b) the variable peak wavelength and (c) the extension to longer wavelength.  Considering these in turn:

(a) The wavelength at half peak intensity ($\lambda_{1/2}$) on the short-wavelength side of the 11.2$\,\mu$m feature in 16 spectra of class A$_{11.2}$ sources
is almost invariant, which we determine to be 11.171~(2$\sigma$~=~0.002)$\,\mu$m; the origin of this striking property is not understood. If the
short-wavelength edge is considered to be part of a Lorentzian profile this would correspond to a FWHM of only \emph{c.} 4~cm$^{-1}$ which is narrow
compared with other UIR bands.

\begin{figure}
 \includegraphics[scale=0.7]{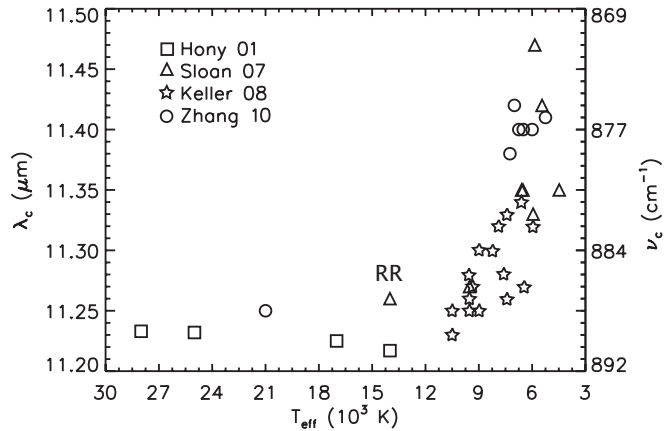}
\caption{Peak wavelength ($\lambda_c$) and class-type of the 11.2$\,\umu$m feature \emph{vs.} effective temperature (T$_{eff}$) of the central star. The squares are a sample of  type A/A(B)$_{11.2}$ sources from
 \protect\cite{Hony01}: GGD~27-ILL (star formation region), CD-42~11721 (Herbig Ae/Be star), NGC~7023 (reflection nebula) and IRAS~21282+5050 (planetary nebula). The triangles are The Red Rectangle (RR) and mostly carbon stars (class B$_{11.2}$ to C$_{11.2}$) from \protect\cite{Sloa07}, the 5-points stars represent Herbig Ae/Be stars (class B$_{11.2}$ to B/C$_{11.2}$) from \protect\cite{Kell08} and the circles are proto-planetary nebulae (class B/C$_{11.2}$  to C$_{11.2}$) from \protect\cite{Zhan10}.}
 \label{11.2_temp}
 \end{figure}

(b) The peak wavelength of the 11.2$\,\umu$m feature has been found to depend on the effective temperature, T$_{\mathrm{eff}}$, of the exciting star
\citep{Sloa07,Kell08}. Fig.~\ref{11.2_temp} collects together data from various authors.  The peak wavelength of the 11.2~$\umu$m band
is independent of temperature for class A$_{11.2}$ objects (squares), but starting from class B$_{11.2}$ ( triangle - The Red Rectangle) it shifts to longer wavelength with decrease in stellar temperature. This change has been interpreted in terms of relatively newly formed material near to low temperature carbon-rich or Herbig Ae/Be stars (class C$_{11.2}$) compared with older processed material in the higher temperature environments of planetary nebulae and
H\,\textsc{ii} regions \citep{Sloa07}.

(c) \cite{Bark87} suggested that the long red tail of the 11.2$\,\umu$m band may be due to superimposed transitions between higher vibrational levels such as
v~=~2~$\rightarrow$~1. In the absence of laboratory data this contribution was incorporated assuming that the v~=~2~$\rightarrow$~1 band falls at an arbitrarily chosen wavenumber 5~cm$^{-1}$ lower than for the fundamental transition. More recently, \cite{Vand04} noted that while a feature of class
A(B)$_{11.2}$ could be considered as coming from the same PAH population distribution as that which gives rise to class A$_{11.2}$ -- but with higher
internal energy, they commented that the difference between the A$_{11.2}$ and B$_{11.2}$ classes could not readily be explained by anharmonic effects
alone. A contribution to emission in the red tail could potentially come from duo-Hs of compact PAH cations \citep{Hudg99}, C--H OOP transitions in PAH
anions \citep{Baus09} or from protonated PAHs \citep{Knor09}. In addition \cite{Wada03} have discussed the possible role of $^{13}$C isotopic substitution. Blind spectral decomposition studies indicate that a contribution to the red wing of `11.2$\,\umu$m' emission arises in carriers well
removed from the exciting star in \emph{e.g.} NGC~7023; this emission has been ascribed to `Very Small Grains' (VSGs) \citep{Rapa06,Bern07,Rose11, Rose12}.

The shapes of the AIBs have been modelled \citep{Schu93,Cook98,Vers01,Path08}; the most detailed study of the 11.2$\,\mu$m band is that of \cite{Pech02}
and includes both intermode anharmonicity (which from laboratory measurements of \cite{Jobl95} allows the effect of cooling on emission wavelength and
linewidth to be estimated), and intramode anharmonicity which is a contribution from vibrational transitions of the type $v$~$\rightarrow$~$v-1$ with
$v$~$\geq$~2. A mass distribution for the PAHs was included of the form $\emph{N}_C^{-3.5}$ where $\emph{N}_C$ is the number of carbon atoms.  This affects
the response of the PAHs to optical excitation and their subsequent cooling.  However, in considering molecular diversity, it was assumed that all PAHs
contributing to the 11.2$\,\mu$m band profile had exactly the same infrared active mode frequency.   It was deduced that the asymmetric profile appeared
to be characteristic of the anharmonicity of molecular modes, and that including a spread of frequencies according to molecular size and geometry would
lead to a change in overall band shape and consequent difficulty in achieving a good fit to the observed band.  This contrasts with the work reported here
where we find that the dependence of the vibrational transition frequencies on the masses of the 11.2$\,\mu$m carriers and their relative abundance are key factors in determining the overall 11.2$\,\mu$m band profile and its variation between objects.

\section{DFT Calculations}
\label{sec:calc}

\subsection{Theoretical Approach}
A large number of DFT calculations of PAH vibrational frequencies have been reported. Attention has generally been focused on neutral and singly ionised
(\emph{i.e.} radical cation) PAH molecules, with extension to hydrogenated, protonated, irregular and other closed-shell charged PAHs \citep{Hudg01,
Beeg01, Path05, Path06, Baus08, Baus09, Baus10, Rick09, Hamm11,Ricc11, Ricc12,Boer14, cand14}.  The work reported here is in two closely related parts:

First, we report the results of DFT calculations on a set of medium-sized PAHs, inspired in part by the work of \cite{Path05} on a series of acenes up to heptacene (see their fig. 7 which shows convergence of the solo OOP wavelength at high mass). We have undertaken a systematic investigation to explore how the frequency and intensity of the out-of-plane (OOP) bending solo modes of neutral PAHs vary with   PAH size and shape. The calculations were performed using Q--Chem \citep{Shao06} with the B3LYP functional \citep{Beck93, Step94} and the 6--31G* basis set. Secondly, DFT calculations were carried out for selected highly symmetric compact large solo-containing PAHs and these results applied in the full spectral profile modelling. Gas-phase laboratory data were used to obtain mode-specific vibrational scaling factors.

 \begin{figure}
 \centering
\includegraphics[scale=0.5]{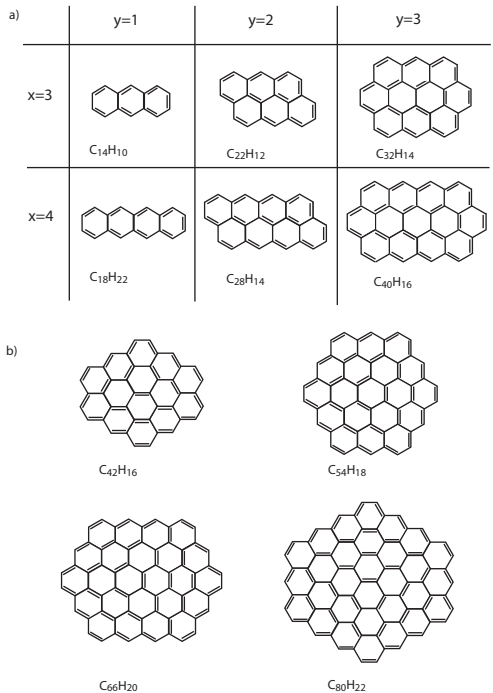}
 \caption{Molecular structures (a) for  PAHs with $x$~=~3-4 and $y$~=~1-3, where $x$ is the number of six-membered rings in a row and $y$ is the number of
 rows. For $y$~=~3, $x$ indicates the number of edge rings in the top and bottom rows.  In (b) four large PAH structures with D${_{2h}}$ and D${_{6h}}$ symmetry are shown.}
 \label{mol_struc}
 \end{figure}

\begin{table}
\begin{center}
\caption{Computed scaled DFT (B3LYP/6-31G*) C--H out-of-plane (OOP) vibrational wavenumbers for (a) acenes with $x$~=~3-8, $y$~=~1 and
multi-row PAHs with $x$~=~3-7, $y$~=~2 and 3 (sf=0.987, see text), and for (b) large compact PAHs (sf=0.975, see text) shown in Fig. \ref{mol_struc}.}
 \begin{tabular}{cccccc}
 \hline
 \hline
 $(x,y)$ & Molecule & Wavenumber & Wavelength & Intensity\\
               &                  &   (cm$^{-1}$)  & ($\,\umu$m)  & (km~mol$^{-1}$)\\
\hline
(a) \\
(3,1) & C$_{14}$H$_{10}$ & 876.7& 11.410 &  55.0 \\
(4,1) & C$_{18}$H$_{12}$ & 895.3& 11.169&  69.3 \\
(5,1) & C$_{22}$H$_{14}$ & 901.7& 11.090 &  87.8 \\
(6,1) & C$_{26}$H$_{16}$ & 905.7 & 11.041 & 106.7 \\
(7,1) & C$_{30}$H$_{18}$ & 906.6 & 11.030 & 124.8 \\
(8,1) & C$_{34}$H$_{20}$ & 906.6 & 11.030 & 146.2 \\
\\
(3,2) & C$_{22}$H$_{12}$ & 875.2 & 11.426 &  96.9 \\
(4,2) & C$_{28}$H$_{14}$ & 881.8 & 11.340 & 109.3 \\
(5,2) & C$_{34}$H$_{16}$ & 886.5 & 11.280 & 123.4 \\
(6,2) & C$_{40}$H$_{18}$ & 888.1 & 11.300 & 133.7 \\
(7,2) & C$_{46}$H$_{20}$ & 888.1 & 11.300 & 172.0 \\
\\
(3,3) & C$_{32}$H$_{14}$ & 884.2 & 11.310 & 123.0 \\
(4,3) & C$_{40}$H$_{16}$ & 890.5 & 11.230 & 155.4 \\
(5,3) & C$_{48}$H$_{18}$ & 891.3 & 11.220 & 185.0 \\
(6,3) & C$_{56}$H$_{20}$ & 891.3 & 11.220 & 204.6 \\
(7,3) & C$_{64}$H$_{22}$ & 891.3 & 11.220 & 231.8 \\
\hline
 (b)\\
    &   C$_{32}$H$_{14}$ & 877.0 & 11.403 & 120.3 \\
    &   C$_{42}$H$_{16}$ & 882.2 & 11.328 & 154.3 \\
 &   C$_{54}$H$_{18}$ & 885.0 & 11.300 & 190.6 \\
 &   C$_{66}$H$_{20}$ & 887.9 & 11.262 & 203.61 \\
 &   C$_{80}$H$_{22}$ & 889.3 & 11.245 & 234.31 \\
 &   C$_{96}$H$_{24}$ & 891.1 & 11.222 & 265.16\\
\hline
\hline
\end{tabular}
\end{center}
\end{table}

In the first part of the study PAH structures are labelled ($x$, $y$) where $x$ is the number of 6-membered rings in a row and $y$ is the number of rows.
Examples include anthracene (C$_{14}$H$_{10}$) for which $x$~=~3 and $y$~=~1 and anthanthrene (C$_{22}$H$_{12}$) for which $x$~=~3 and $y$~=~2
(see Fig.\,\ref{mol_struc}(a) and Tab.\,1(a)). Vibrational frequencies computed in DFT calculations are systematically high and a scaling factor is often invoked to bring calculation and experiment into agreement. Here a scaling factor for the specific solo OOP mode was employed. For $y$~=~1, 2 and 3 PAHs a scaling factor of 0.987 was adopted in the modelling, determined by comparing the high-resolution gas-phase experimental value for anthracene of 876.7~cm$^{-1}$ \citep{Cane97} with the unscaled DFT result of 888.2~cm$^{-1}$.

The second part of our study focussed on large compact PAHs (see Fig.\,\ref{mol_struc}(b) and Tab.\,1(b)). These molecules were chosen to investigate whether the OOP frequency shares a similar mass-dependent behaviour as for low-medium mass PAHs.  The 11.2$\,\umu$m band is generally attributed to large neutral PAHs \citep{Hony01, Baus08, Ricc12}, their compactness and high symmetry making them optimal candidates to survive in harsh astronomical environments. The study was again undertaken with B3LYP/6-31G*, using Gaussian~03 \citep{g03}.  A scaling factor of 0.975 was used for these molecules, deduced by reference to the experimental gas-phase data for the solo OOP bending mode of ovalene (C$_{32}$H$_{14}$) \citep{Jobl94}.

The use of a scaling factor is normal practice when comparing laboratory and DFT results. Typically, a scaling factor of 0.958 for B3LYP/4-31G  is used  for simplicity for the entire spectrum, based on comparison with matrix-isolation data. However, it is known that this technique introduces unpredictable shifts in the band positions \citep{Lang96}. In contrast to previous studies, the scaling factor employed in this work is referenced to gas-phase data and is specific to the vibrational mode (solo OOP transitions of PAHs).  For our analysis, this is preferable to the use of a generic scaling factor.  At the present time there is insufficient laboratory gas-phase data on a range of PAHs to evaluate the applicability of the same scaling factor for all PAHs. However, the use of the well characterised  B3LYP/6-31G*  and the mode-specific scaling factor should reduce the uncertainty significantly.

\begin{figure}
 \centering
 \includegraphics[scale=0.7]{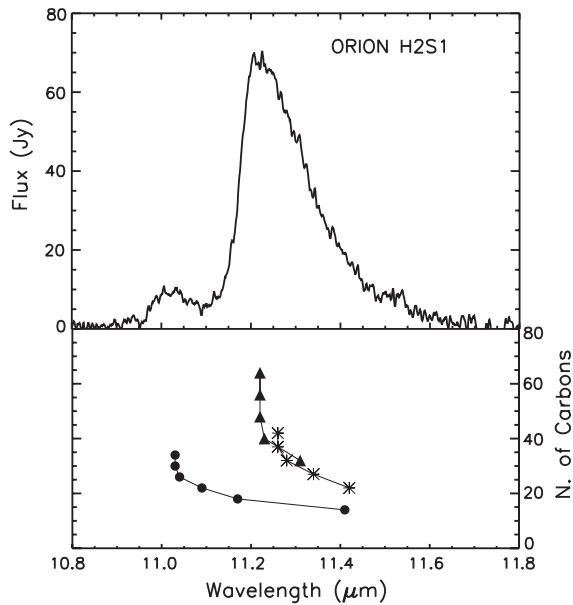}
 \caption{Upper panel: ISO-SWS spectrum of the 11.0 and 11.2$\,\umu$m band complex in Orion~H2S1, an example of class A$_{11.2}$. Lower panel: wavelength
 behaviour of the OOP bending transition with respect to number of carbon atoms for $y$~=~1 (filled circles - acenes), $y$~=~2 (stars) and $y$~=~3 (filled triangles) - see Fig.\,\ref{mol_struc}(a) and Tab.\,1(a). }
 \label{comp}
 \end{figure}

\subsection{Results \& Discussion}
\label{sec:res disc}

Tab.\,1 collects together the computed intensities and unscaled vibrational wavenumbers of the studied molecules. Fig.\,\ref{comp} shows that for the acenes ($x=\,3-8$, $y\,=\,1$) the wavelength decreases with increase in $x$ (or number of carbon atoms), converging towards a high-mass asymptotic value of 11.03$\,\umu$m.  A similar characteristic holds for zig-zag edge molecules with two rows (stars) and three rows (filled triangles), but they reach the high-mass limiting value at longer wavelength (Fig.~\ref{comp}, lower panel). In all cases, the wavelength for the solo OOP bending mode moves to shorter wavelength as the number of carbon atoms (\emph{i.e.} the mass) increases (see Tab.\,1 (a) and (b)).  A similar dependence of the CH OOP frequency on PAH mass is found in the computed data of \citet{Baus08} who used the B3LYP functional and a (smaller) 4-31G basis set; the values for C$_{54}$H$_{18}$ and C$_{96}$H$_{24}$ are 11.061 and 10.959$\,\umu$m, respectively.

Larger PAH molecules with $\mathrm{N_C}\,>\,100$ are needed to determine the value at the asymptotic limit but this was beyond our computational capacity. Hence a functional form was used to fit the available data (once scaled) from Tab.\,1(b) and so deduce the asymptotic wavelength limit for the largest PAHs: $\omega=(b+a/\mathrm{N_C}^2)^{-1}$, where $\omega$ is the mode wavenumber and $\mathrm{N_C}$ the number of carbon atoms. The parameters of the best fitting are $a=0.021\,\pm0.002$ and the asymptotic value $b=11.2139\,\pm\,0.0007\,\umu$m (or $891.75\,\pm\,0.06$~cm$^{-1}$). Thus, the inferred asymptotic limit for large PAHs falls to shorter wavelength compared with low-medium mass PAHs, i.e. PAHs with multiple rows  with $y$ ~=~ 2 and 3 as shown in Fig.~3(a). The value of 11.21$\,\umu$m falls within the range of peak wavelengths (11.20 to 11.24$\,\umu$m) in class \textbf{A$_{11.2}$} objects \citep{Vand04} and falls very close to the observed value of $\lambda_{1/2}$ of 11.171\,$\umu$m discussed in Section 2.2(a).

Key results from these calculations are a) the more extended (acene) and larger PAH molecules have solo out-of-plane modes at the shortest wavelengths, b) the transition wavelengths for acenes and for large PAHs reach asymptotic values at high mass, c) increasing the number of rows (larger PAHs) results in a faster approach to the asymptotic limit than for the single-row acenes, d) the higher-mass acenes may contribute to 11.0$\,\umu$m emission.  The presence of a high-mass asymptote for the wavelength of the largest PAHs provides an attractive explanation for the steepness and wavelength invariance of the short-wavelength side ($\lambda_{1/2}$) among class \textbf{A$_{11.2}$} objects. In the following sections we explore this further using a detailed theoretical model.

\section{Emission Model}
\label{sec:model}
In this section we describe how the DFT results for frequencies and intensities are incorporated into a model for the AIB emission occurring in
astrophysical environments. As far as we are aware this is the first attempt to include mass-dependence of the PAH emission wavelength into a model and
using, where available, reference to experimental results.  We have chosen to employ a model based on the thermal approximation \citep[as employed by \emph{e.g.}][]{Pech02,Path08}.

\subsection{Theoretical Approach}
\label{sec:theoapp}

\cite{Lege89} proposed that the thermal approximation can be used to describe the infrared cooling of PAHs where this treatment assumes that a molecule can
be considered as a heat bath with an average molecular energy $U$ and temperature $T$. Following absorption
of a UV photon, a PAH molecule has an internal energy $U(T_{\mathrm{p}})$, where $T_{\mathrm{p}}$ is the (initial) peak temperature.

In the harmonic approximation, the internal energy can be written:
 $$U(T)= \sum_{i=1}^{n} \frac{hc\,\omega_i}{\exp(hc\,\omega_i / k_{\mathrm{B}}T)-1}$$
 where $\omega_i$ is the wavenumber of the $i$th vibrational mode and $n$ is the total number of modes of the molecule. The PAH molecule cools through its various vibrational modes, through a so-called \emph{radiative cascade}. For the $i$th mode the emission rate $\phi_i$ is given by:
 $$\phi_i= \frac{A_i^{1,0}}{\exp(hc\,\omega_i/ k_{\mathrm{B}}T)-1}$$
  where the Einstein coefficients $A_i^{1,0}$ for spontaneous emission can be calculated from the DFT transition intensities \citep{Cook98}.

 The fractional energy emitted in the $i$th mode, corresponding to a fall in internal energy $\delta U$, is:
 $$\delta E_i(T)=\frac{\phi_i \times{\omega_i}}{\sum_{i=1}^n \phi_i \times{\omega_i}}\,\delta U(T)$$

 The total emitted energy is obtained by integration over the temperature range from $T_{\mathrm{p}}$ to 50~$K$, and weighted by the rate of photon absorption
 $$R_{abs}=\int^{13.6}_0 \frac{B^d_{\nu} \sigma_{\nu}}{h\nu}d\nu$$

 where $\nu$ is the frequency of the absorbed photon, $\sigma_{\nu}$ is the frequency-dependent photo-absorption cross-section and 13.6~eV represents the high-energy cut-off in the radiation field.  For each molecule $\sigma_{\nu}$ was taken from the French-Italian database\footnote{\texttt{http://astrochemistry.oa-cagliari.inaf.it/database}.} \citep{Mall07}.  For a molecule (PAH1) for which
 there was no entry, the cross-section was estimated using data available for the molecule closest in size (PAH2), scaled by
 $N_{\mathrm{C}}^{\mathrm{PAH1}}/N_{\mathrm{C}}^{\mathrm{PAH2}}$ \citep{mul06}. The excitation source is represented by a diluted Planck function $B^d_{\nu}$
 (see Tab.\,\ref{tab_2}).
At high photon energy, photoionisation of the neutral PAH can also be significant and leads to a reduction in the quantum yield for IR emission. To take  this process into account, we incorporated in our modelling the energy-dependent photo-yield using the empirical law given by \cite{Lepa01} and ionisation energies (IE) from the French-Italian database. When experimental values were not available, a theoretical vertical IE was used.

\begin{table}
\begin{center}
\caption{Physical conditions for objects considered in the modelling. T$_{\mathrm{eff}}$ represents the effective temperature of the central star in K, L the luminosity in units of solar luminosity (L$_{\sun}$), W$_{\mathrm{dil}}$ the geometric dilution factor and D the distance in kpc.  In the case of the HD 44179, the two components of the binary stellar system are taken into account.}
 \begin{tabular}{cccc@{\hspace{4pt}}c}
 \hline
 Object                               & Orion\,H2S1             & NGC\,7027                   & \multicolumn{2}{c}{HD\,44179}\\
 \hline
  Class$^{(a)}$                     &  A$_{11.2}$        & A(B)$_{11.2}$           & \multicolumn{2}{c}{ B$_{11.2}$ }\\
 T$_{\mathrm{eff}}$\,(K) & $4\cdot10^{4\,b}$   & $1.6\cdot10^{5\,d}$  &         8250$^f$      & $6\cdot10^{4\,f}$ \\
 L (L$_{\sun}$)                & $1.35\cdot10^{7\,c}$ & 10$^e$                         &  $6\cdot10^{3\,f}$&                100$^f$\\
 W$_{\mathrm{dil}}$      & $2.09\cdot10^{-11}$ & $2.8\cdot10^{-17}$    & $4.8\cdot10^{-6}$ & $3.7\cdot10^{-13}$\\
 D (kpc)                              & 0.45$^b$                  & 1$^e$                      &  \multicolumn{2}{c}{0.71$^f$} \\
 \hline
 \end{tabular}
\label{tab_2}
\end{center}
{\bf References}:   (a) \citealt{Vand04},  (b) \citealt{Odel01}, (c) \citealt{Vers01}, (d) \citealt{Bein96}, (e) \citealt{Buja01}, (f) \citealt{men02}.
\end{table}

\subsection{Experimental input}
\label{sec:expin}

 \citealt{Jobl95} studied the influence of temperature on  IR absorption spectra of gas-phase PAHs in the laboratory and found a linear dependence on
 temperature for both band position and band width, arising mostly from anharmonic coupling between modes. The following relations for the
 frequency $\omega(T)$, and width $\Delta\omega(T)$, were used in the modelling:

 \begin{equation}\label{eq:T}
   \omega(T) = \omega_0 +\Delta\omega_{RS} + \chi' T
 \end{equation}

 and
\begin{equation}
 \Delta\omega(T)=\Delta\omega(0)+\chi'' T
\end{equation}

where $\Delta\omega_{RS}= \omega_L(0)-\omega_0$ can be considered to be an empirical red-shift (RS) between the calculated DFT frequency (at zero Kelvin) $\omega(0)$ and the frequency $\omega_L(0)$, at zero Kelvin inferred  from the experimental (laboratory) studies  \citep{Cook98}. Tab.\,\ref{tab:coef} collects the coefficients for the solo out-of-plane mode. We consider ovalene and anthracene to be representative of large and small (acene) PAH molecules, respectively. In the case of ovalene, $\chi''$ was derived after removing the rotational contribution (see Fig.\,8 of \citealt{Pech02}).  For a molecule such as ovalene and based on experiments in
absorption, it is expected that the emission will move to longer wavelength (lower $\omega$) as the cooling cascade occurs, and that the linewidth will
increase.

\begin{table}
\begin{center}
\caption{Empirical coefficients for the temperature-dependent wavenumber and linewidth of the solo OOP mode (see equations (1) and (2).}
 \begin{tabular}{l|cc}
 \hline
                                           &   Large$^{\dag}$     & Acenes$^{\ddag}$ \\
 \hline
  $\chi'$ (cm$^{-1}$/T)                             &   -0.0114  & -0.0065 \\
  $\Delta\omega_{RS}$  (cm$^{-1}$)&  -12.91     & -9.54 \\
  \hline
  $\chi''$    (cm$^{-1}$/T)                          &  0.0157  &  --   \\
  $\Delta\omega(0)$  (cm$^{-1}$)      &   0.54       &   --  \\
  \hline
\end{tabular}
\label{tab:coef}
\end{center}
{\bf Note}:  $^{\dag}$Measurements on neutral gas-phase ovalene in the 550-820~K range and in a Ne matrix at 4~K  \citep{Jobl94,Jobl95,Pech02}.\\
 $^{\ddag}$  Measurements on neutral gas-phase anthracene \protect\citep{Semm91,Cali62,Cane97} and in an Ar matrix at 10\,K \protect\citep{Hudg98}. No information is available on the temperature dependence of the linewidth; a value for anthracene in the gas phase (at 370~K) is known \citep{Cane97}.
\end{table}

\subsection{Single molecule calculation: influence of temperature?}
\label{sec:sing mol}

\begin{figure}
 \centering
 \includegraphics[scale=0.934782609]{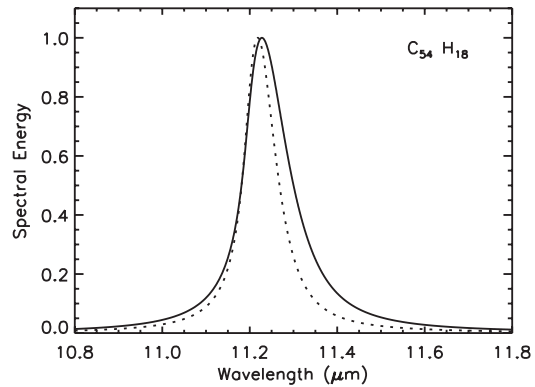}
 \caption{Emission profile computed for C$_{54}$H$_{18}$ exposed to stellar radiation with $T_{eff}$ = 8,250~K (dotted line) and 160,000~K (full line). The
 linewidth is taken to be constant with a FWHM of 5~cm$^{-1}$.}
 \label{sing mol}
 \end{figure}

We consider here the emission profile resulting from the cooling cascade for a single PAH molecule,  circumcoronene (C$_{54}$H$_{18}$, Fig.3 b), in radiation fields with T$_{eff}$=~8,250~K and 160,000~K corresponding to HD~44179 and NGC~7027, respectively. The results are given in Fig.\,\ref{sing mol} and show that (a) the peak
wavelength is red-shifted and (b) the redward side extends to longer wavelength for a PAH with a high initial internal temperature (\emph{i.e.} irradiation
as in NGC~7027). However, this is not in agreement with observation. The observed peak wavelength in NGC~7027 is lower and the red wing is no more extended
than in HD~44179 (see Fig.\,\ref{em_mod}). We also remark that a red tail originating from `hot-band' emission from higher vibrational levels is difficult
to reconcile with the presence of Class A$_{11.2}$ features (with a low tail-to-top ratio) in many compact and extended H\,$\textsc{ii}$ regions
\citep{Vand04}.

We conclude that a factor other than the radiation temperature is responsible for the change in 11.2$\,\umu$m profile between objects and, given the DFT results of Section 3., propose
that the dependence of the emission wavelength on PAH mass is the largest single factor in determining the differences in the 11.2$\,\umu$m profiles
between objects.  This is now explored through modelling of astronomical emission profiles which includes the wavelength dependence on PAH mass.

\subsection{Summary of calculations}
\label{sec:sum_calc}
 The procedure to calculate the energy E(T) emitted in a band is described in Sec.~\ref{sec:theoapp}. Each E(T) is then associated with a Lorentzian
 profile so that the total energy emitted in the solo out-of-plane bending mode is described by:
 $$
 E_{solo}(\omega)= \sum_m \alpha_m \sum_T \frac{(\Gamma/2) E_m(T)}{\pi\,((\Gamma/2)^2+(\omega-\omega_m(T))^2)}
 $$
 where $\Gamma=\Delta\omega_m(T)$ is the FWHM, $\alpha_m$ is a multiplicative coefficient and $m$ is summed over the molecules considered. This calculation
 is then weighted by $R_{abs}$ and a $\chi^2-$minimisation routine employed to determine the best fit to the astronomical ISO spectra of the three sources
 - Orion H2S1, NGC~7027 and HD~44179 (see Tab.\,\ref{tab_2}) - with the smallest possible number of contributing molecules. The only variable in the $\chi^2-$minimisation is $\alpha_m$ which is the number of molecules of type $m$ included in the fitting.

\section{Astronomical profile Modelling}
 \label{sec:res}

In this section we describe the results of modelling the 11.2$\,\umu$m band for three profile classes: (A$_{11.2}$, A(B)$_{11.2}$ and B$_{11.2}$). It was found that only four very symmetrical solo-containing molecules (C$_{32}$H$_{14}$, C$_{42}$H$_{16}$, C$_{54}$H$_{18}$ and C$_{66}$H$_{20}$, Tab.1~b) were needed to produce acceptable fits, the sole variable in the $\chi$$^2$ optimisation being the relative proportions of these PAHs ($\alpha_m$) which are taken to be representative of the PAH mass distribution.  However, we emphasise that this does not represent identification of these particular four PAHs; rather the results show that solo-containing neutral PAHs in this absolute and relative size range provide a good fit to the 11.2$\,\umu$m feature and its profile variation with object. This aspect is discussed further in Section 6.1. Two closely related models were developed.

 \subsection{Model 1: 11.2$\,\umu$m modelling using temperature-dependent linewidths}
\label{sec:temdepmodel}

\begin{figure}
 \centering
\includegraphics[scale=0.75]{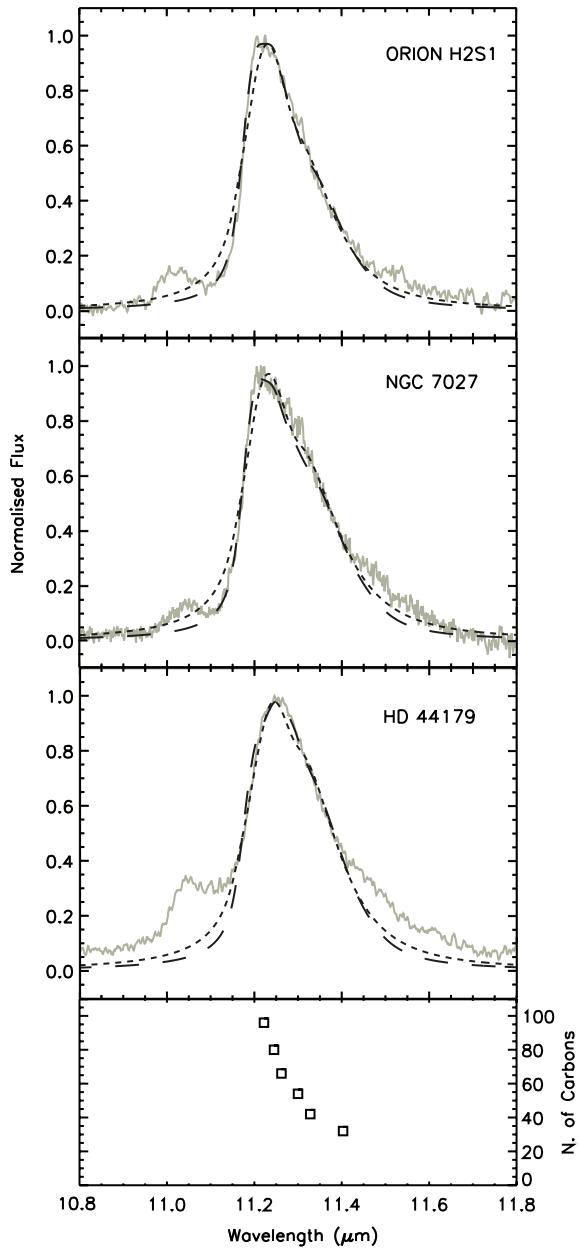}
 \caption{Continuum subtracted ISO-SWS spectra of objects of class A$_{11.2}$ - Orion~H2S1, A(B)$_{11.2}$ - NGC~7027 and B$_{11.2}$ - HD~44179, the Red
 Rectangle (grey line), superimposed with the emission model results using (Model 1) temperature-dependent $\Delta\omega(T)$ (short dashed line) and (Model 2) molecule-specific constrained widths
 $\Delta\omega$ (long dashed line) - see text for details. The lower panel shows the (scaled) wavelength behaviour of the solo OOP mode of large PAHs from Table.1 b.}
 \label{em_mod}
 \end{figure}

In the first application of the above approach ( short-dashed line in Fig.\,\ref{em_mod}), the experimentally-based temperature-dependence for the band frequency and width (Eq.~1. and 2. respectively - see Section 4.2) for ovalene were taken to hold for the larger molecules in the set (\emph{i.e.} independent of PAH mass). However, as each molecule has a different initial peak temperature (determined by the stellar SED) and a different vibrational solo-OOP frequency according to molecular size, the cooling cascade and resultant emission profile necessarily differs in each case. A $\chi^2$ minimisation was undertaken with the only variable being the relative adundance of the four PAHs. This model provides a reasonable match with the peak wavelength and long-wavelength tails of the bands, but it does not account very well for the steep short-wavelength side seen in many astronomical spectra including Orion~H2S1 and NGC~7027.  This discrepancy falls in the part of the profile where high-mass PAHs are expected to contribute most strongly and lies near to the asymptotic limit discussed in Section~\ref{sec:res disc}.\\

Considering the experimental observations for the widths of the infrared active C--H stretch transitions in naphthalene, pyrene and coronene\footnote{Ovalene was not considered because the law was deduced from only two experimental points.} (Equation\,2), then for a given temperature the width of the band is found to decrease with increase in PAH mass \citep[see Tab.\,2 and Fig.\,4 in][]{Jobl95}; for example at 600~K the linewidth values for naphthalene, pyrene and coronene are 40, 26 and 15~cm$^{-1}$, respectively. The solo-OOP transition linewidth behaves similarly; at 370~K the experimental solo-OOP linewidths in absorption for anthracene \citep{Cane97} and ovalene (Joblin, C., personal communication) are 25~cm$^{-1}$ and 6~cm$^{-1}$, respectively. Additionally, as higher mass PAHs do not reach as high an internal temperature -- for the same exciting SED, high-mass PAHs have narrower absorption (and emission) band widths than their lower mass counterparts. This aspect is addressed in Model 2.

\subsection{Model 2: molecule-dependent temperature-independent linewidths}

 Assuming that a Lorentzian profile describes the short wavelength side of the 11.2$\,\umu$m band for Class A$_{11.2}$ and A(B)$_{11.2}$ objects, then from  profile fitting of the short-wavelength part of the profile we obtain a Lorentzian peak wavelength of 11.198$\,\umu$m (or 893.02\,cm$^{-1}$) and a FWHM of only 0.05$\,\umu$m (or 4\,cm$^{-1}$).  This FWHM represents the upper limit on the width of a single contribution to the shorter wavelength side of the 11.2$\,\umu$m band.  A  similar value has been suggested by \cite{Cami11}. Using this information, Fig.\,\ref{em_mod}  shows the best $\chi^2-$ minimised fit with constrained FWHM values of 10, 7.5, 5 and 4~cm$^{-1}$ for C$_{32}$H$_{14}$, C$_{42}$H$_{16}$, C$_{54}$H$_{18}$ and C$_{66}$H$_{20}$, respectively. The value of 10~cm$^{-1}$ was taken for ovalene with reference to the laboratory absorption data (at 600~K) and the 4~cm$^{-1}$ value as constrained by the observational data (see above and  Section~\ref{sec:temdepmodel}).
 The fit to the steep short-wavelength side of the 11.2$\,\umu$m band for the class A$_{11.2}$ and A(B)$_{11.2}$ objects is improved, while the long-wavelength tail does not change significantly. Thus an accurate description of the shorter wavelength side of the 11.2$\,\umu$m band requires lower linewidths than experimentally found for the FWHM of \emph{e.g.} ovalene.  We also note the Red Rectangle, and in part NGC~7027, has a discrepancy in the 11.45-11.65$\,\umu$m region in both models.  We return to this in Section~\ref{sec:11.4-11.7region}.

\subsection{Effect of SED on computed 11.2$\,\umu$m profiles}
\label{sec:effect of SED}

In section~\ref{sec:sing mol} it was shown that for a single PAH molecule, C$_{54}$H$_{18}$, the calculated emission profile with a low
T$_{eff}$ SED is narrower than for higher T$_{eff}$. To explore the effect of the excitation SED on computed 11.2$\,\umu$m profiles, we have taken the (fixed) mass distribution which was determined for NGC~7027 using the SED for NGC~7027, and calculated the profile using a lower T$_{eff}$ SED - that of the Red Rectangle. The calculated result in Figure 7. (upper panel, full line) is qualitatively similar to the single-molecule
calculation and gives a very poor match to the astronomical spectrum for NGC~7027 when compared with Figure 6. middle panel, Model 2.  Similarly, taking the (fixed) mass distribution as deduced for the Red Rectangle, using a high-temperature SED (that of NGC~7027) also yields a poor match (Figure 7. lower panel, full line).  Clearly the SED used in the calculation plays a role in determining a computed 11.2$\,\umu$m profile but the SED is fixed for a given object and the observed profile is determined principally by the solo-containing PAH mass distribution.

  \begin{figure}
 \centering
\includegraphics[scale=0.7]{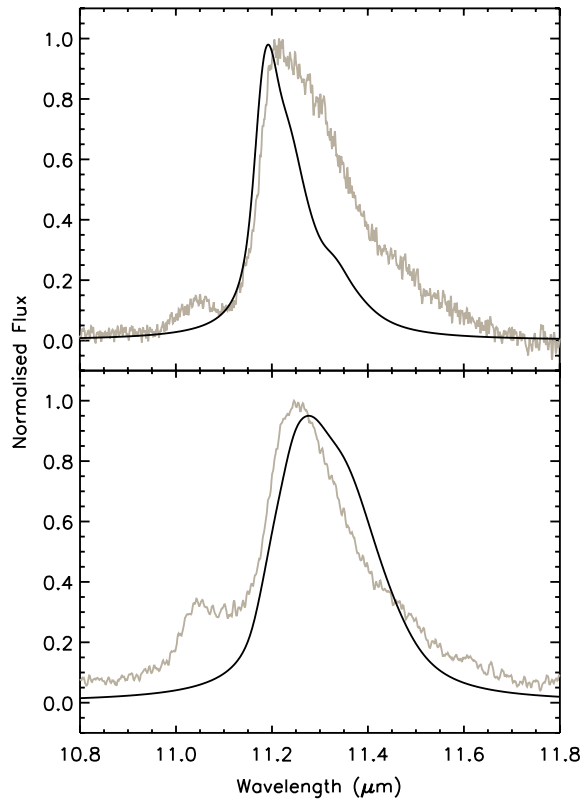}
 \caption{Upper Panel. Calculated results for the 11.2$\,\umu$m in the spectrum of NGC~7027 (grey) using the mass distribution inferred for this object from Model 2 and using the SED for the Red Rectangle. Lower Panel. Calculated results for the astronomical 11.2~$\mu$m  band in the Red Rectangle (grey) using the mass distribution inferred for this object from Model 2 and using the SED of NGC~7027.}
\label{em_7027 SED}
\end{figure}

\section{Analysis and Discussion}
\label{sec:dis}

\subsection{PAH mass distribution and the 11.2$\,\umu$m profile}
\label{sec:massdistbn}

 Taking the sum of the contributions of the two lower mass PAHs (C$_{32}$H$_{14}$ and C$_{42}$H$_{16}$) to the 11.2$\,\umu$m band fit in Model 2, and similarly taking C$_{54}$H$_{18}$ and C$_{66}$H$_{20}$ as representative of higher mass PAHs, an indicative low mass:high mass abundance ratio can be determined with the break falling at N$_C\sim$50. Using the $\chi^2$ fitted results, this ratio is 1.8:1 for Orion~H2S1, 2.4:1 for NGC~7027 and 3.7:1 for the Red Rectangle. From this we infer that the Red Rectangle PAH population comprises lower mass PAHs than for NGC~7027 and Orion~H2S1. We deduce that the shape and peak wavelength of the 11.2$\,\umu$m feature can be described principally in terms of the PAH mass distribution, where high-mass PAHs are responsible for the steep short wavelength side. The low variance of $\lambda_{1/2}$ and the steep blue side in A$_{11.2}$ objects arise as a natural consequence of the asymptotic limit reached by the solo OOP mode in high-mass PAHs. In  class B$_{11.2}$ objects such as the Red Rectangle, the less steep short-wavelength side, the more extended redward tail and a band maximum at longer wavelength suggests that the PAH mass distribution is skewed towards lower masses than in class A$_{11.2}$ objects.  The class A(B)$_{11.2}$ object NGC~7027 then falls between the class A and B extremes with a broad spread of PAH masses as suggested by the intermediate value of the mass distribution indicator introduced above.
Using \emph{Spitzer} observations of NGC~7023, \citet{Boer13} fitted the PAH emission spectra as a function of offset from the star
HD~200775 using the NASA Ames PAH IR Spectroscopic Database (PAHdb) \citep{Baus10}.  The shape of the 11.2$\,\umu$m feature evolves from class A$_{11.2}$ relatively near the star (\emph{c.} 20$''$ offset) towards B$_{11.2}$ as a denser region is approached at higher offset. It was
concluded that the overall PAH spectra result from approximately equal proportions of large and small PAHs at Position I (a dense region) with
larger PAHs being more prominent in the diffuse region (Position II).  Our interpretation of the 11.2$\,\umu$m band shape is consistent with these results, though our approach differs in that our inferred mass distributions hold specifically for neutral solo-containing PAHs.

 \begin{figure}
 \centering
\includegraphics[scale=0.75]{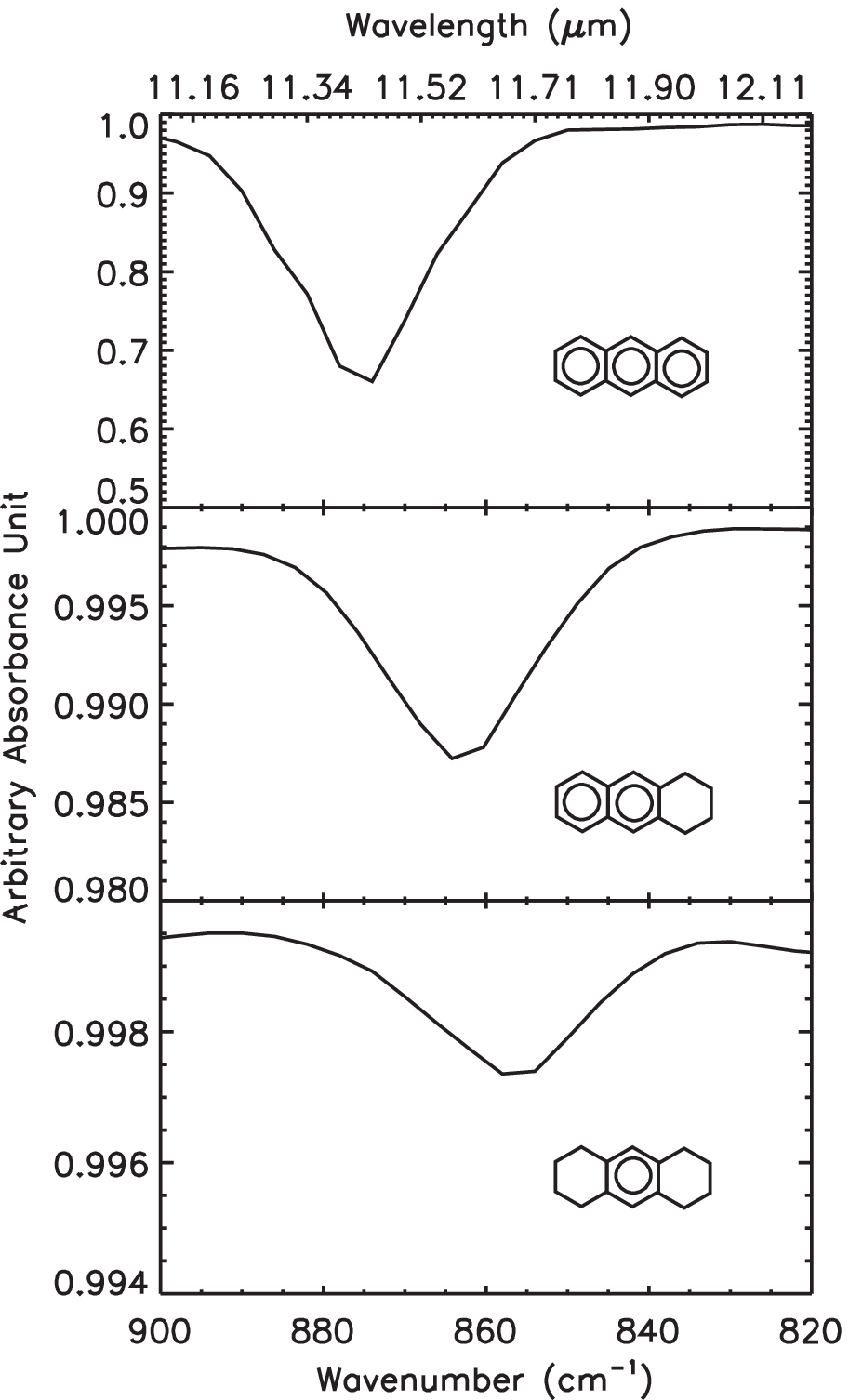}
 \caption{Experimental gas-phase absorption spectra (T$\sim500$~K) of anthracene (C$_{14}$H$_{10}$, top panel), 1,2,3,4-tetrahydroanthracene
 (C$_{14}$H$_{14}$, middle panel) and 1,2,3,4,5,6,7,8-octahydroanthracene (C$_{14}$H$_{18}$, bottom panel). The data are taken from \protect\cite{Stei11}.
 As the number of attached hydrogens increases, the peak position of the solo C-H out-of-plane bending mode moves to longer wavelength.}
 \label{hpah}
 \end{figure}

\subsection{Emission in the 11.4-11.7$\,\umu$m region}
\label{sec:11.4-11.7region}

The Red Rectangle and to a lesser extent Orion~H2S1 and NGC~7027, have excess emission relative to the models in the 11.4-11.7$\,\umu$m wavelength region.  Based on the results of DFT calculations and the gas-phase experimental OOP value for anthracene of 11.4$\,\umu$m (at 373~K, see Fig.6), it is suggested the excess in the Red Rectangle is due to emission from relatively low-mass PAHs which were not included in the modelling of the 11.2$\,\umu$m feature presented here.  The scaled DFT out-of-plane bending modes for anthracene C$_{14}$H$_{10}$ ($x=3$; $y= 1$) and anthanthrene C$_{22}$H$_{12}$ ($x = 3$; $y = 2$) fall at 11.4$\,\umu$m.  This too is consistent with a PAH population of lower mass for the Red Rectangle deduced separately in section 6.1.  Noting the usual redward shift that occurs in emission, these and similar molecules probably contribute to emission in the 11.4-11.7$\,\umu$m region.

A second possible contribution to the `excess' could arise from hydrogenated PAHs. Addition of hydrogen atoms to PAH molecules (H-PAHs) perturbs the C-H out-of-plane bending mode frequency, thus affecting the peak position. Fig.\,\ref{hpah} shows gas-phase absorption spectra in the 11.4$\,\umu$m region, recorded at $\sim$500 K \citep{Stei11} for the closely related molecules: anthracene (C$_{14}$H$_{10}$), 1,2,3,4-tetrahydroanthracene (C$_{14}$H$_{14}$) and 1,2,3,4,5,6,7,8-octahydroanthracene (C$_{14}$H$_{18}$). Starting with anthracene (upper panel), increasing the number of hydrogen atoms (middle and lower panels) moves the peak position for the solo C-H out-of-plane bending mode (Hs in the 9,10 positions) to longer wavelength. In a major study of infrared spectra of hydrogenated PAHs in an argon matrix it was found that aromatic C-H stretching bands near 3.3$\,\umu$m weaken and are replaced with stronger aliphatic bands near 3.4$\,\umu$m, and that aromatic C-H out-of-plane bending mode bands in the 11-15$\,\umu$m region shift and weaken \citep{Sand13}.  However, very few PAHs in the sample set had solo-hydrogens.  We tentatively suggest that a signal identified by \cite{Rose11} could be due to solo-containing hydrogenated PAH molecules rather than `Very Small Grains' - VSGs (see their Fig.\,6); further assessment of this will be described elsewhere.  Significantly, objects classified as A(B)$_{11.2}$ and B$_{11.2}$, show aliphatic bands at 3.4 and 3.52$\,\umu$m \citep{Geba85} as would be expected to appear when PAH hydrogenation levels are high; these bands appear in the experimental spectra of hydrogenated anthracene, but emission in the 6-9$\,\umu$m region remains weak as for other neutral PAHs.

\subsection{The 11-15$\,\umu$m and wider spectral region}
\label{sec:11-15region}

\begin{figure}
 \centering
 \includegraphics[scale=0.75]{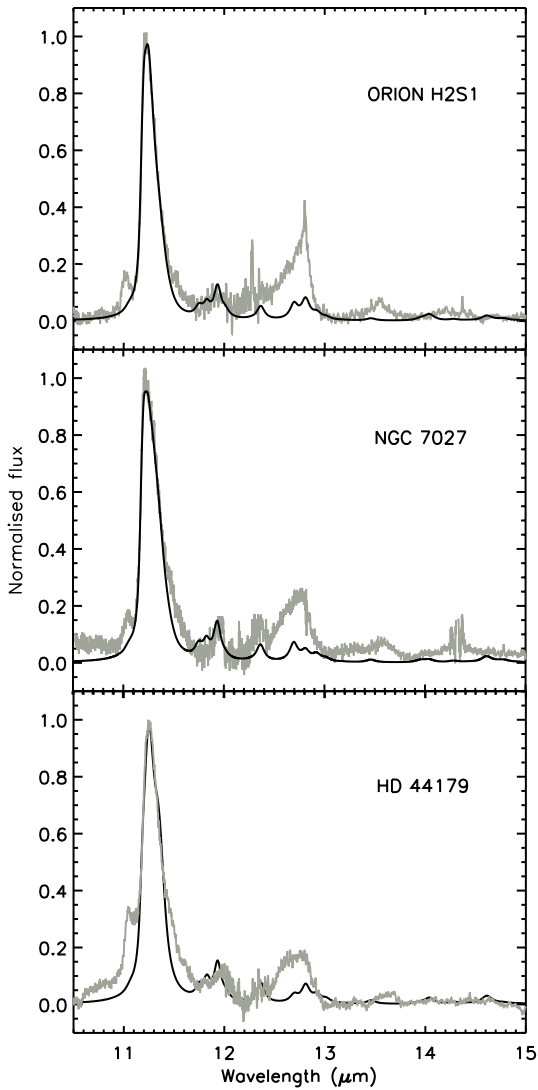}
 \caption{Emission model in the fingerprint region for the astronomical objects as in Fig.~\ref{em_mod}. A scaling factor of 0.979 and a fixed FWHM of
 12~cm$^{-1}$ were applied. To combine the fit for the 11.0-11.2$\,\umu$m band complex and the remaining bands in the fingerprint region, the molecular
 contribution for the solo mode in a PAH  was assumed to be a Lorentzian function peaking at the scaled (sf$\,=0.979$) DFT frequency and with FWHM of
 10~cm$^{-1}$. }
 \label{em_mod_fing}
 \end{figure}

 The molecules considered here give a strong contribution to the UIR bands in the 11-15$\,\umu$m range \citep{Schu93}. Fig.\,\ref{em_mod_fing} shows the  fingerprint region for the three prototype objects compared with the emission model results. Due to the lack of T-dependent experimental data for the position of and width of the bands in the fingerprint region, a single scaling factor of 0.979 and a fixed FWHM of 12~cm$^{-1}$ were applied for all bands. To combine the fit for the 11.0-11.2$\,\umu$m band region and the remaining bands in the fingerprint region, we assumed that the molecular contribution for the solo mode in a PAH can be modelled as a Lorentzian function peaking at the scaled (sf$\,=0.979$) DFT frequency but with a FWHM of 10~cm$^{-1}$. The predictions for the bands in the fingerprint region are consistent with the astronomical
 spectra of the three objects, with no more than a small contribution to the 12.7$\,\umu$m feature. The model slightly overestimates the flux of the 12.0$\,\umu$m band for the Red Rectangle, which is generally ascribed to duo hydrogen modes \citep{Hony01}.  However, in SWS-ISO data, the region around 12$\,\umu$m is sensitive to the way spectra are combined, thus introducing extra uncertainty of 20-30 percent \citep{Hony01}. A further issue is placement of the continuum in the subtraction applied to the spectra. The lack of a detailed treatment of anharmonicity (temperature dependence) may also affect this modelled spectrum. It is notable that the model built on treatment of the 11.2$\,\umu$m region does not introduce features which are not present in astrophysical spectra of the 11-15$\,\umu$m region; this is also the case for the 6-11$\,\umu$m region. For example, the strongest calculated vibration of C$_{54}$H$_{18}$ in the 6-9$\,\umu$m region has a strength of ~50 km~mol$^{-1}$ \citep{CandT} compared with 191 km~mol$^{-1}$ for the solo OOP mode, i.e. an intensity ratio of 0.3. The effect of the radiative cascade during the PAH emission will reduce this ratio even more.  Also, the 6-9$\,\umu$m region is generally attributed to ionised PAHs, which show very strong C-C and C-H in plane stretching modes, compared to their neutral counterparts \citep{Hudg99}. As for the 3.3$\,\umu$m band, it is known that DFT calculations overestimate the intensity of the C-H stretching modes \citep{Lang96}, making it difficult to directly compare our model with the astronomical spectra.

 \subsection{Emission in the 11.0$\,\umu$m region}
 \label{sec:11.0 discussion}

 \subsubsection{Emission in the 11.0$\,\umu$\MakeLowercase{m} region - a summary}
Given the proximity of 11.0~$\umu$m emission to the 11.2~$\umu$m band, we review here whether there is significant evidence for a related origin. \cite{Roch91} first identified a faint feature on the short-wavelength side of the 11.2~$\umu$m band which appears quite prominently in spectra of a  number of objects \citep{Hony01}. The most commonly occurring band has an approximately symmetric shape centered at 11.05 $\umu$m, a rather narrow FWHM~of~0.1 $\umu$m and an intensity which ranges between 2 and 10\% of the 11.2~$\umu$m feature. In a few cases a somewhat higher 11.0/11.2 ratio is found: two Herbig Ae/Be stars, MWC~1080 \citep{Sako07} and HD~37411\citep{Boer08}, in the  H\,\textsc{ii} region IRAS~18434-0242 \citep{Roel96} and in two reflection nebulae, NGC~1333~SVS3 \citep{Sloa99} and NGC~7023 \citep{Wern04, Boer13}.

 Emission features in the 11.0$\,\umu$m region have been described as `blue outliers' of the 11.2~$\umu$m band by \cite{Sloa99}. On increasing distance from the hot star SVS3 it was found that a shorter wavelength part at $\sim$~10.8~$\umu$m disappeared first, whereas $\sim$~11.0~$\umu$m emission was more persistent. Using the results of theoretical calculations by \cite{Lang96}, \cite{Sloa99} considered whether the `11.0~$\,\umu$m' emission might be due to neutral acenes, such as anthracene or pentacene, but following comparison with IR matrix data they concluded that this was unlikely and PAH cations were preferred as carriers. \cite{Hudg99} discussed this in some detail and extended consideration across the 11-14~$\,\umu$m region. The idea of a cationic origin gained support from DFT calculations on large solo-containing PAHs \citep{Baus08, Ricc12} where the (scaled) solo wavelengths for the  cations were found to fall to wavelengths shorter than that for neutrals. \cite{Wern04} and \cite{Sako07} found that the \emph{c.} 11.0~$\umu$m emission reduces in intensity moving away from the exciting star in NGC~7023 and MWC~1080, respectively, a result for NGC~7023 recently reinforced by \emph{Spitzer} observations \citep{Boer13}.

 The most commonly discussed carrier of the \emph{c.}~11.0~$\umu$m emission carrier is PAH cations.  However, \cite{Povi07} observed a lack  of variation in the shape and/or intensity of the 11.0~$\umu$m band in the photodissociation region M17-SW, even where the level of the ionisation in the gas varied with position. Moreover, in recent work on the evolution of the 11.2~$\umu$m and 11.0~$\umu$m features SE of the Orion Bright Bar \citep{Boer12}, the integrated strength of the main 11.2~$\umu$m feature dropped by a factor of 5.7 between 2'.6 and 5'.7 whereas the 11.0$\,\umu$m feature dropped by a factor of 2.7. This result was described as `counterintuitive' in terms of the 11.0$\,\umu$m emission being due to PAH cations but rationalised as being influenced by the degree of hydrogenation.  In a recent \emph{Spitzer} study of the mid-infrared structure of the massive star-formation region W49A further puzzles have emerged; the 11.0/11.2 ratio is found to be constant across the whole region and with a very low value of $\sim$0.01 \citep{Stoc14}. The ratio is more commonly in the range 0.02-0.2 as discussed by \cite{Stoc14}.  In a recent study fitting the spectrum of NGC~7023 with the NASA AMES PAH database, \cite{Boer13} noted that the 11.0~$\umu$m band could be reproduced by ionised nitrogen-containing PAHs. In summary there are many questions as to the origin(s) of feature(s) in the 11.0~$\umu$m region.

  \subsubsection{Red Rectangle emission in the 11.0$\,\umu$\MakeLowercase{m} region}

\begin{figure}
 \centering
  \includegraphics[scale=0.75]{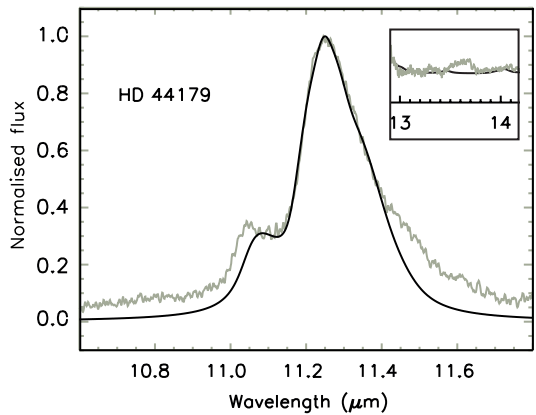}
  \caption{Comparison between modelled and astronomical spectrum of the Red Rectangle (grey line) for the 11.0-11.2 $\,\umu$m band
 complex. The fit takes into consideration the four large PAHs plus pentacene, hexacene and heptacene. The width for the contribution is fixed at 8\,cm$^{-1}$ for acenes and as in Model 2 for large PAHs. The value of the low mass: high mass ratio of 3.7: 1 deduced in section 6.1 is changed by less than 10\% on including acenes in the mode. The insert shows a comparison between the contribution of quarto OOP bending modes in acenes and the Red Rectangle spectrum (grey line) aroun 13.5~$\umu$m. The contribution is fitted using the constraints described in Sec \ref{sec:11-15region}.}
 \label{em_110_112}
 \end{figure}

It was inferred earlier that the Red Rectangle has a population of lower-mass PAHs than for Orion~H2S1 and NGC~7027 (Section~\ref{sec:massdistbn}), so it is of interest to explore whether low-mass PAHs including acenes might contribute to the 11.0$\,\umu$m region in this object.  It is notable that DFT calculations predict an asymptotic wavelength for the longer acenes which falls  at  a shorter wavelength than the 11.2~$\umu$m band (Section~\ref{sec:res disc}).  We have conducted a Model~2-type calculation including acenes which yields the result shown in Fig.~\ref{em_110_112}.  The fit to the 11.0$\,\umu$m region is improved and the 11.2$\,\umu$m profile is not affected, suggesting that, at least for this object, the acene series may make a significant contribution to 11.0$\,\umu$m emission. 
The lack of a very good fit between the 11.0$\,\umu$m feature and larger acenes may be due to the underestimation of the frequency of the OOP solo mode transition for these molecules.  We are exploring this further by undertaking DFT calculations which include anharmonicity explicitly.

 The IR vibrational spectrum of neutral, long acenes is dominated by strong C-H oop modes (solo and quarto) \citep{Path05}. We therefore checked the contribution of the acene quarto modes by modelling them (Model 2) and compared the result with the emission in the Red Rectangle (Fig.\ref{em_110_112}, small insert). Clearly, the prediction is consistent with the astronomical spectrum. The overall result that the Red Rectangle has a population of relatively low mass PAHs may be of significance in determining the origin of the strong unidentified optical emission bands seen in this object \citep{Warr81}.

\section{Conclusions}
Based on the results of DFT calculations, we created a detailed emission model and applied it to a small set of PAH molecules with the aim of
understanding profile  and profile variations in the 11.2$\,\umu$m band.

The model shows that the 11.2$\,\umu$m feature, regardless of the A/B classification, can be fitted by a mass distribution of a few, large neutral PAHs
containing solo hydrogens. We infer that the mass distribution is principally responsible for the red tail of the feature and the steep short-wavelength side, and hence offers a good explanation for the 11.2$\,\umu$m band asymmetry and its variation between and within objects. In this picture, large neutral solo-containing PAHs dominate class A$_{11.2}$ objects and smaller PAHs class B$_{11.2}$ objects, whereas class A(B)$_{11.2}$ possesses a more even mass distribution. We also note that a PAH mass range of 30-70 carbon atoms compares favourably with the masses of fullerenes, recently discovered in several objects (\emph{e.g.} \citealt{Cami10, Sell10}). The present model gives a more detailed description of the different physical effects (anharmonicity, photo-ionisation and astronomical environment) involved in building up the feature.

In this paper we described the results of a detailed emission model applied to the solo mode in PAHs and compared them to astronomical data. Based on the results we find that the profile of the 11.2$\,\umu$m band and its variation are well described principally in terms of the mass distribution of neutral solo-containing PAHs.

\section*{Acknowledgments}
We thank Dr. Amit Pathak for useful discussions, Dr. Christine Joblin for unpublished information on the temperature dependence of the ovalene spectrum, and Dr. Jesse Bregman for very helpful comments. AC acknowledges STFC for a studentship and The University of Nottingham and the Royal Astronomical Society for financial support. PJS thanks the Leverhulme Trust for award of a Research Fellowship and the Leiden Observatory for hospitality that allowed completion of this work.  
All quantum chemical calculations were performed using the University of Nottingham High Performance Computing Facility. ISO is an ESA project with instruments funded by ESA Member States (especially the PI countries: France, Germany, The Netherlands, and the UK) and with the participation of ISAS and NASA.

\label{lastpage}

\end{document}